# Entropy production in a mesoscopic chemical reaction system with oscillatory and excitable dynamics

Ting Rao, Tiejun Xiao, ZhonghuaiHou[1]

Hefei National Lab for Physical Sciences at Microscale & Department of Chemical Physics, University of Science and Technology of China, Hefei, Anhui, 230026, P.R. China

**Abstract:** Stochastic thermodynamics of chemical reaction systems has recently gained much attention. In the present paper, we consider such an issue for a system with both oscillatory and excitable dynamics, using catalytic oxidation of carbon monoxide on the surface of platinum crystal as an example. Starting from the chemical Langevin equations, we are able to calculate the stochastic entropy production $P$ along a random trajectory in the concentration state space. Particular attention is paid to the dependence of the time averaged entropy production $P$ on the system size $N$ in a parameter region close to the deterministic Hopf bifurcation.In the large system size (weak noise) limit, we find that $P \sim N^\beta$ with $\beta = 0$ or $1$ when the system is below or above the Hopf bifurcation, respectively. In the small system size (strong noise) limit, $P$ always increases linearly with $N$ regardless of the bifurcation parameter. More interestingly, $P$ could even reacha maximum for some intermediate system size in a parameter region where the corresponding deterministic system shows steady state or small amplitude oscillation. The maximum value of $P$ decreases as the system parameter approaches the so-called CANARD point where the maximum disappears. This phenomenon could be qualitatively understood by partitioning the total entropy production into the contributions of spikes and of small amplitude oscillations.

## 1. Introduction

Very recently, stochastic thermodynamics (ST) has gained considerable attention, ever since the pioneer work of Udo Seifert[1-7]. ST provides a framework for describing small systems like colloids or biomolecules driven out of equilibrium but still in contact with a heat bath. A first law like energy balance involving exchanged heat and work done, as well as entropy production entering the refinements of second law, can be defined consistently along a stochastic trajectory. Importantly, some general fluctuation relations hold for these stochastic trajectory-based thermodynamic variables. The total entropy change, $\Delta s_{tot}$, is shown to be related to the dynamical irreversibility of the trajectory. It also obeys the so-called integral fluctuation theorem (FT), $\langle \exp(-\Delta s_{tot}) \rangle = 1$, which gives the second law $\langle \Delta s_{tot} \rangle \geq 0$. In nonequilibrium steady states, a detailed form of FT also holds for $\Delta s_{tot}$, i.e., $p(\Delta s_{tot})/p(-\Delta s_{tot}) = e^{\Delta s_{tot}}$, where $p(\cdot)$ denotes the probability that $\Delta s_{tot}$ takes a certain value. Albeit ST was originally applied to Brownian particles described by over-damped Langevin equation, the concepts and principles can also be applied to general stochastic systems described by master equations. Although first law-like energy balance should be properly interpreted for general stochastic dynamic systems, trajectory-based entropy production can be well-defined, such that second-law and FT can be well-established. Such approaches may lead us closer towards a systematic understanding of non-equilibrium statistical mechanics of small systems, in general. In this context, ST has found wide applications in optically driven colloids[4], (bio)chemically driven enzymes[7], state transition in biomolecules[7], general chemical reaction networks[6], to list just a few. It is worthy to note that another kind of framework provided by P. Gaspard et al. [8-12] for describing nonequilibrium thermodynamics and dynamic irreversibility in small systems based on master equation description has also been successfully applied to out-of-equilibrium nanosystems, including chemical reactions.

---

[1]To whom correspondence should be addressed. Email address: hzhlj@ustc.edu.cn

In two recent papers, we have applied ST to mesoscopic chemical oscillation systems with supercritical Hopf bifurcations (HB) [13-14]. Our main motivation is to unravel the interplay between the nonlinear dynamic behaviors far from equilibrium and ST features. We first considered the conceptual Brusselator model, described by a master equation. Although the reactions are irreversible, the transitions in the molecular number state space are reversible, which serves as a kind of micro-reversibility based on which time-reversed trajectory can be well-defined. We have calculated the entropy production $P$ along a stochastic limit cycle by using Gillespie algorithm, in a parameter region close to the HB. Interestingly, $P$ shows distinct scaling behaviors with the system size V at different side of the HB: It increases linearly with V in the deterministic oscillatory region, while it is independent with V in the steady state region. Such an observation implies that one may use ST features to characterize the occurrence of a HB, which can be viewed as a kind of nonequibrium phase transition, in mesoscopic chemical systems. To further demonstrate this, we later extended our study to a general reaction network, described by chemical Langevin equations. Herein, the reversibility is related to the continuous concentration space and requires only at a coarse-grained level, much more relaxed than that in the molecular-number space. This time, the entropy production along a stochastic trajectory can be calculated by using a path-integral approach. Thanks to the stochastic normal form theory we developed before to account for noise-induced oscillation and coherence resonance [15-16], we were able to obtain the analytical expression for $P$ when the system is close to HB. The analysis clearly showed that $P$ scales as $V^{\alpha}$ in the limit $V \to \infty$, where $\alpha$ equals to 0, 0.5, or 1 when the system below, at, or above the HB, respectively. These two studies shed some new lights on the application of ST in characterization of typical nonlinear dynamic features, here the oscillation associated with HB.

In the present paper, we continue these series of study to a mesoscopic chemical reaction system with both oscillatory and excitable dynamics. Excitability is of ubiquitous importance in many physical, chemical and biological systems. It describes any stable dynamical system that exhibits pulses when the amplitude of a perturbation exceeds a fixed threshold. In spatially extended systems, this leads to the generation of fronts and spiral waves, most notably occurring on surface catalytic reactions on single crystal surfaces and on human heart leading to fibrillation[17-22]. Excitability usually involves time-scale separation in the dynamic evolutions of a fast and a slow variable. In case when both oscillatory and excitable dynamics coexist in a system, two types of oscillation can be observed: One is quasi-harmonic oscillation with small amplitude growing from the HB, the other is relaxation oscillation with large amplitude and multi-time scales. In such systems, small oscillations may change abruptly to large spikes in an exponentially narrow parameter regime, known as CANARD phenomenon. Following our previous studies, we are thus wondering how such nontrivial dynamic features would influence the ST properties, especially the scaling behaviors of the entropy production.

To this end, we have applied the concept of ST to a model system, carbon monoxide (CO) oxidation on platinum (Pt) surface. We choose this system for two-fold reasons. On one hand, very abundant nonlinear dynamic behaviors have been observed in this system, including multistability, oscillation, excitability, as well as CANARD phenomenon. On the other hand, CO oxidation systems are of both theoretical and experimental interest, and our study may help understand the nonequilibrium fluctuation properties of this important system. We consider that the surface is divided into many mesoscopic cells, whose space scale is determined by the diffusion length. The reactions inside each cell are considered to be homogeneous, while concentration gradient may exist between neighboring cells.

For simplicity and as the first step, we only consider a single cell in the present work. Extension to spatially extended systems might be possible but is beyond the scope of the present study.

We start from the chemical Langevin equation (CLE) describing the dynamics inside a cell containing N lattice sites. By using path integral approach, expressions for the entropy production $P$ can be obtained, which depend on the detailed dynamics along astochastic trajectory. We focus on a parameter region close to the deterministic Hopf bifurcation, and particular attention is paid to the dependence of $P$ on the system size $N$, here denoting the number of lattice sites inside a mesoscopic surface cell. Our numerical results show that P always increases linearly with $N$ when $N$ is small, regardless of the control parameter. In the large $N$ limit, however, $P$ may scale as $N^1$ or $N^0$ depending on the parameter value. In the intermediate range of $N$, $P$ can even show a maximum. By partitioning $P$ into two parts, one contributed from the spikes and the other from small amplitude oscillations, and investigating the dependences of each part on $N$ separately, we can then qualitatively illustrate the scaling laws of $P$ and the occurrence of the maximum.

## 2. Model and Results

The model used in the present paper was developed for the oxidation of CO on Pt(110) on single crystal surface[23-24]. The reaction follows a Langmuir-Hinshelwood mechanism, which involves the adsorption of CO and $O_2$ molecules, desorption of CO molecule, and the reaction between adsorbed CO molecule and O atom. In addition, the adsorbate-induced $1 \times 1 \Leftrightarrow 1 \times 2$ phase transition is taken into account to address the influence of the surface structure on the reactivity. The state of a cell containing $N$ adsorption sites can be described by $X_N(t) = [N_{CO}(t), N_O(t), N_{1 \times 1}(t)]^T$, where $N_{CO}$, $N_O$, and $N_{1 \times 1}$ denote the number of adsorbed CO molecules, oxygen atoms, and adsorption sites in a non-reconstructed $(1 \times 1)$ surface, respectively. According to these mechanisms, there are six reaction channels as listed in Table 1, where we have used $\mathbf{x} = (u, v, w)$ to stand for the concentrations of $(N_{CO}(t), N_O(t), N_{1 \times 1}(t))$. Note that the transition rates $a_{i=1,\ldots,6}$ are all proportional to the system size $N$.

According to the stochastic processes and transition rates shown in Table I, the CLE for the current model reads

$$\frac{du}{dt} = \frac{1}{N}\left[(a_1 - a_3 - a_4) + \sqrt{a_1}\eta_1(t) - \sqrt{a_3}\eta_3(t) - \sqrt{a_4}\eta_4(t)\right],$$

$$\frac{dv}{dt} = \frac{1}{N}[(2a_2 - a_4) + 2\sqrt{a_2}\eta_2(t) - \sqrt{a_4}\eta_4(t)], \tag{1}$$

$$\frac{dw}{dt} = \frac{1}{N}[(a_5 - a_6) + \sqrt{a_5}\eta_5(t) - \sqrt{a_6}\eta_6(t)].$$

where $\eta_{i=1,\ldots,6}(t)$ are Gaussian white noises with $\langle \eta_i(t) \rangle = 0$ and $\langle \eta_i(t)\eta_j(t') \rangle = \delta_{ij}\delta(t-t')$. The items with $\eta_i(t)$ give the internal noises, which scale as $1/\sqrt{N}$ because $a_{i=1,\ldots,6} \propto N$. In the macroscopic limit $N \to \infty$, the internal noise items can be ignored and the system's dynamics is described by the deterministic equation,

$$\dot{x}_j = \frac{1}{N}\sum_{\varrho=1}^{M} v_\varrho^j a_\varrho(\mathbf{x}) \equiv f_j(\mathbf{x})\ (x_{j=1,2,3} = u, v, w) \tag{2}$$

where $\mathbf{v}_1 = (1,0,-1,-1,0,0)^T$, $\mathbf{v}_2 = (0,2,0,-1,0,0)^T$, $\mathbf{v}_3 = (0,0,0,0,1,-1)^T$ and $M = 6$ is the number of reaction channels.

TABLE I. Stochastic processes and reaction rates for *CO* oxidation on *Pt(110)*. All the parameter values are the same as those listed in Table 2 of Ref. 23.

| Process | Rate | Descriptions |
|---|---|---|
| $N_{CO} \to N_{CO} + 1$ | $a_1 = NP_{CO}k_{CO}S_{CO}(1-u^\xi)$ | CO adsorption |
| $N_O \to N_O + 2$ | $a_2 = \frac{1}{2}NP_O k_O[S_O^{1\times 2}(1-w) + S_O^{1\times 1}w](1-u)^2(1-v)^2$ | $O_2$ adsorption |
| $N_{CO} \to N_{CO} - 1$ | $a_3 = N[k_{des}^{1\times 2}(1-w) + k_{des}^{1\times 1}w] \times u$ | CO desorption |
| $N_{CO} \to N_{CO} - 1, N_O \to N_O - 1$ | $a_4 = Nk_{re}uv$ | Reaction |
| $N_{1\times 1} \to N_{1\times 1} + 1$ | $a_5 = Nk_{1\times 1}(1-w) \times f_{1\times 1}(u,w),$ with $f_{1\times 1}(u,w) = (1-\varepsilon)u^\lambda + \varepsilon w^\lambda$ | (1×2) to (1×1) |
| $N_{1\times 1} \to N_{1\times 1} - 1$ | $a_6 = Nk_{1\times 2}w \times f_{1\times 2}(u,w),$ with $f_{1\times 2}(u,w) = (1-\varepsilon)(1-u)^\lambda + \varepsilon(1-w)^\lambda$ | (1×1) to (1×2) |

The Fokker-Planck equation (FPE) corresponding to Eq. (1) reads

$$\frac{\partial p(\mathbf{x};t)}{\partial t} = -\sum_i \frac{\partial}{\partial x_i}[f_i(\mathbf{x})p(\mathbf{x};t)] + \frac{1}{2N}\sum_{i,j} \frac{\partial^2}{\partial x_i \partial x_j}[G_{ij}(\mathbf{x})p(\mathbf{x};t)] \tag{3}$$

where $p(\mathbf{x};t)$ is the probability distribution in the concentration space, $f_j(\mathbf{x})$ is the deterministic term defined in Eq.(2), and $G_{ij}(\mathbf{x}) = 1/N \sum_{\varrho=1}^{M} v_\varrho^i v_\varrho^j a_\varrho(\mathbf{x})$. By introducing probability current density

$$J_i(\mathbf{x}) = \frac{1}{2}\sum_j G_{ij}\left(H_j - \frac{1}{N}\frac{\partial}{\partial x_j}\right)p(\mathbf{x};t) \qquad (4)$$

where $H_j = 2\sum_k \Gamma_{jk}\tilde{f}_k$ with $\sum_j G_{ij}\Gamma_{jk} = \delta_{ik}$ and $\tilde{f}_k = f_k - 1/(2N)\sum_j(\partial G_{kj})/(\partial x_j)$, we can write the FPE in a compact form $\partial_t p(x;t) = -\sum_i(\partial J_i/\partial x_i)$.

With the variation of the parameters $P_O$ and $P_{CO}$, the system (2) shows very abundant bifurcation features [23-24]. In the present paper, we fix $P_O = 9.6 \times 10^{-5}$ mbar, $T = 520K$ and choose $P_{CO}$ as the only control parameter. In such a case, the bifurcation diagram of the deterministic system (2) is shown in Fig. 1. There is a supercritical Hopf bifurcation (*HB*) at $P_{CO} \approx 3.557 \times 10^{-5}$, where a stable limit cycle emerges. The limit cycle disappears via a saddle-node infinite period (*SNIPER*) bifurcation when it encounters the turning point (*TP1*) at $P_{CO} \approx 3.6151 \times 10^{-5}$. For $P_{CO}$ less than *HB* or larger than *TP1*, the system only shows one stable state. Stable oscillations can only be observed in the region between *HB* and *TP1*. In addition, an interesting feature of the system is the existence of a CANARD point at $P_{CO} \approx 3.575 \times 10^{-5}$, where a very fast transition from a small amplitude oscillation to a large amplitude oscillation occurs [25].

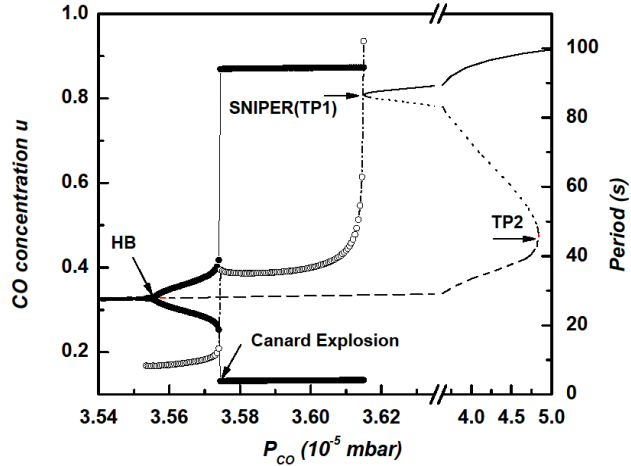

FIG.1. Bifurcation diagram for the deterministic system (2). HB stands for the supercritical Hopf bifurcation at $P_{CO} \approx 3.557 \times 10^{-5}$, TP1 and TP2 denote two turning points at $P_{CO} \approx 3.6151 \times 10^{-5}$ and $P_{CO} \approx 4.839 \times 10^{-5}$, respectively. Stable limit cycles exist in the region between HB and TP1, where the solid circles show the concentration range of the oscillations. Note the oscillation ends at TP1 via a saddle-node infinite period (SNIPER) bifurcation. The heavy solid lines denote stable steady states, the dashed line unstable states, and the dotted line saddle states. The dependence of the oscillation period between HB and TP1 is depicted by the open circles (the right axis). Importantly, there is a CANARD point at $P_{CO} \approx 3.575 \times 10^{-5}$. The bifurcation diagram is calculated by use of the BIFPACK software (Ref. 26).

In Fig. 2, typical stochastic oscillations for four different system sizes are shown. For $N = 10^8$, the internal noise is rather small and the dynamics of system approximately obeys the deterministic equation (2), hence the $CO$ concentration only shows slight fluctuations around the deterministic steady state as shown in the bottom panel. When the system size decreases to a smaller value, e.g. $N = 10^6$, random fluctuation around thesteady state changes to 'stochastic oscillation' with small amplitude. With further decreasing of the system size, we find that occasional random pulses are triggered on the background of the small stochastic oscillations. For an optimal cell size, $N = 10^4$ for instance, the pulse can become rather regular.

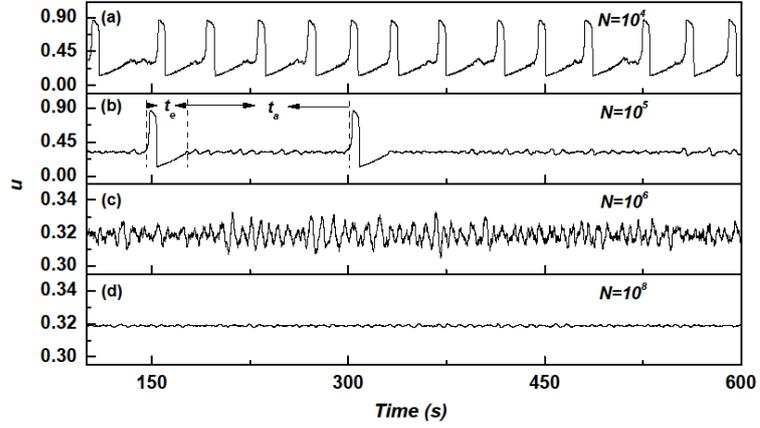

FIG.2. Typical time series of CO concentration for different system sizes N for $P_{CO} = 3.55 \times 10^{-5}$ mbar. From top to bottom, N reads $10^4$, $10^5$, $10^6$, and $10^8$, respectively. Note that the vertical axis has different scales in panels (a) and (b) from those in (c) and (d). For relatively small N, the stochastic oscillations are of large amplitude, while for large N, the amplitude is small. For $N \sim 10^5$, neither type of stochastic oscillations dominates. In panel (b), $t_a$ and $t_e$ denote the activation and excursion time period, respectively.

To apply the ST to this excitable system, we now consider a path $\chi(t)$ generated by Eq.(1) starting from $\mathbf{x}_0(\tau = 0)$ selected from some normalized distribution $p_0(\mathbf{x}_0)$ and ending at $\mathbf{x}_t(\tau = t)$ with normalized distribution $p_1(\mathbf{x}_t)$. Correspondingly, the time-reversed path $\tilde{\chi}(t)$ starts from $\tilde{\mathbf{x}}_0 = \mathbf{x}_t$ and ends at $\tilde{\mathbf{x}}_t = \mathbf{x}_0$ with $\tilde{\mathbf{x}}(\tau) = \mathbf{x}(t - \tau)$. According to Seifert[2], one may define the system entropy along this single trajectory as

$$S(\tau) = -\ln p(\mathbf{x}(\tau), \tau) \qquad (5)$$

where $p(\mathbf{x}(\tau), \tau)$ is the solution of the Fokker-Planck equation evaluated along the trajectory at time $\tau$.

As shown in Ref.[9], by evaluating the entropy balance along the trajectory, one may calculate the so-called 'medium' entropy change as follows,

$$\Delta s_m = V \int_0^t dt \sum_i H_i \dot{x}_i, \qquad (6)$$

where $\dot{x}_i$ is evaluated along $\chi(t)$. It can be shown that the total entropy change along the path $\Delta s_{tot} = \Delta s_m + \Delta s$ obeys the second law like inequality $\Delta s_{tot} \geq 0$ and integral FT $\langle e^{-\Delta s_{tot}} \rangle = 1$ [2]. In the stationary state, a detailed FT $p(\Delta s_{tot})/p(-\Delta s_{tot}) = e^{\Delta s_{tot}}$ also holds. Note that for long time trajectories, $\Delta s$ only contributes a boundary term to $\Delta s_{tot}$ and is much smaller than $\Delta s_m$, one expects that both types of FT also hold approximately for $\Delta s_m$. We can use the time-averaged entropy production $P$ to measure the dynamic dissipation rate

$$P \equiv \lim_{t \to \infty} \frac{\langle \Delta s_m \rangle}{t} = V \sum_i \langle\langle H_i \dot{x}_i \rangle\rangle_s \qquad (7)$$

where $\langle\langle \cdot \rangle\rangle_s$ means averaging over both time and the stationary distribution. We will mainly investigate how $P$ depends on the system size $N$ and the control parameter $P_{CO}$.

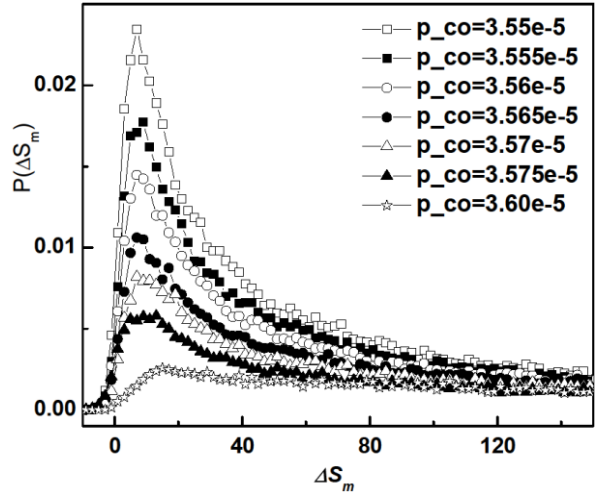

FIG.3. Typical distributions $p(\Delta s_m)$ of the medium entropy change in different parameter regions. System size is N=2×10$^5$.

In this study, $10^5$ stationary trajectories with length $t = 0.5$ are generated to calculate the distribution of medium entropy change $\Delta s_m$ via Eq. (6). The results are plotted in Fig.3 for different $P_{CO}$ with fixed system size $N = 2 \times 10^5$. It is noted that the distributions are strongly non-Gaussian. Apparently the distribution shows no significant difference when the control parameter increases from the steady state region to the oscillation region, and neither the Hopf bifurcation nor the CANARD explosion plays any role here. We note here that the ensemble average of $\Delta s_m$ is quite large here, and the probabilities of paths with negative $\Delta s_m$ are fairly small. If we want to verify the detailed FT $\frac{p(\Delta s_m)}{p(-\Delta s_m)} = e^{\Delta s_m}$ or integral FT $\langle e^{-\Delta s_m} \rangle = 1$ here, much more realizations of $\Delta s_m$ are required.

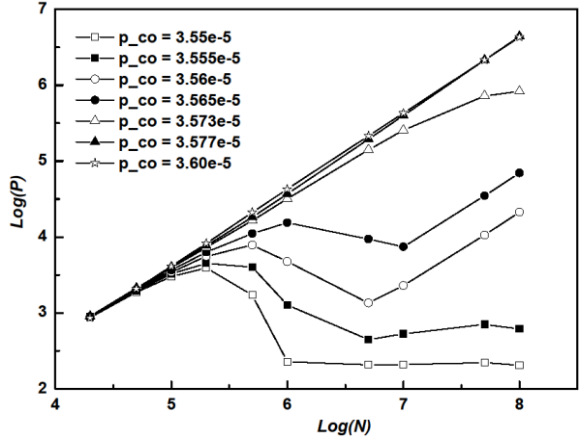

FIG.4. Dependence of the mean entropy production $P$ on the system size $N$ for different $P_{CO}$.

To quantify the dissipation of the system, the mean entropy production $P$ is calculated according to Eq.(7). As shown in Fig.4, we have considered the effects of system size $N$ and control parameter $P_{CO}$. Several interesting features can be observed. For $P_{CO}$ in the steady region, e.g., $P_{CO} = 3.55 \times 10^{-5}$ and $3.555 \times 10^{-5}$, scaling law $P \propto N^0$ is observed in the large system size (weak noise) limit, which is in accordance with the prediction of our previous studies[9]. In addition, it is noted that $P$ could scale as $P \propto N^1$ in the small system size (large noise) limit. More interestingly, $P$ shows a clear-cut maximum at some optimal system size. For $P_{CO}$ between the HB and CANARD where the deterministic system shows small-amplitude oscillation, e.g., $P_{CO} = 3.56 \times 10^{-5}$ and $3.565 \times 10^{-5}$, $P$ scales as $N^1$ for both small and large N, while a maximum still appears at some intermediate N. The scaling for large N is also in agreement with the results in Ref. [9]. For $P_{CO}$ very close to the CANARD, e.g., $P_{CO} = 3.573 \times 10^{-5}$, it seems that the maximum moves to very large N. For $P_{CO}$ to the right side of the CANARD where deterministic large amplitude relaxation oscillations are observed, e.g., $P_{CO} = 3.577 \times 10^{-5}$ and $3.60 \times 10^{-5}$, $P$ scales as $N^1$ in the whole range of N and the maximum disappears. In this relaxation oscillation region, $P$ is not sensitive to the bifurcation parameter $P_{CO}$. We note that such nontrivial behaviors are quite different from the system we studied in Ref.[9], where no excitability exists.

Intuitively, one expects that these features must be relevant to the noise-induced dynamics of such a system.

Although the normal form analysis presented in our previous work [9] can reasonably address the scaling of $P$ for large $N$ for $P_{CO}$ near the HB, it cannot explain the appearance of the maximum and the scaling for $P_{CO}$ in the relaxation region. Here we try to give a qualitative discussion. As shown Fig.2, each stochastic trajectory can be divided into two parts of segments: One consists of the small amplitude oscillations during the activation time $t_a$, the other contains the spikes (relaxation oscillations with large amplitude) during the excursion time $t_e$. If we define the factor $\alpha \equiv \langle t_e \rangle / (\langle t_a \rangle + \langle t_e \rangle)$ as the time percentage of spikes in the trajectory, then one may write,

$$P = [\alpha P_1 + (1-\alpha) P_2], \qquad (8)$$

where $P_1$ and $P_2$ are the mean entropy production corresponding to the spikes and small amplitude oscillations, respectively. We can separately investigate the dependences of $P_1$, $P_2$ and $\alpha$ on the system size $N$, thus may provide an reasonable explanation to the curves in Fig.4.

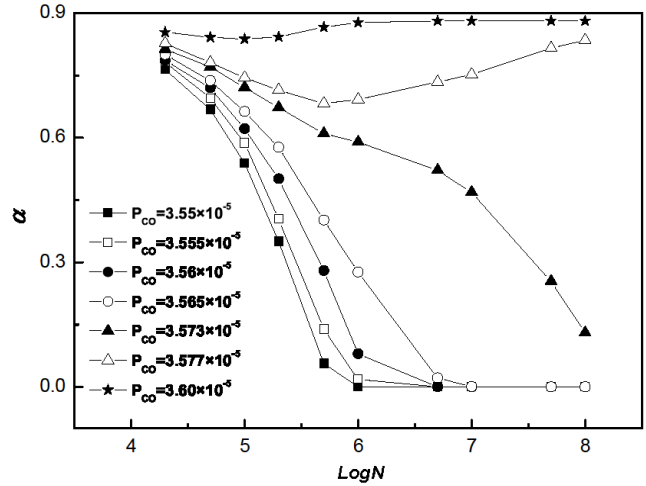

FIG.5. Dependence of the time percentage factor $\alpha$ on system size $N$ for different $P_{CO}$.

In Fig.5, we have plotted the dependence of $\alpha$ on $N$ for different $P_{CO}$ as those in Fig.4. Clearly, two distinct types of tendency can be observed. For $P_{CO}$ less than the CANARD, $\alpha$ decreases quickly to nearly zero as N becomes large. In this region, the system is excitable, and the spike occurs via some type of barrier crossing, thus $\alpha$ is related to the first passage time for such an escape problem. Assume that the barrier height is not dependent on the system size $N$, and notice that the $1/N$ measures the level of internal noise, one may assert that $\alpha \propto e^{-bN}$, where $b$ is a constant not dependent on $N$. However, the fitting is not good for the data shown in Fig.6, indicating that the dynamics is much more complex. For $P_{CO}$ larger than the CANARD, one can see that $\alpha$ is close to 1, and not sensitive to the system size N. This is reasonable because the deterministic system already shows spikes in this parameter region. In Fig.6, $P_1$ and $P_2$ are presented as functions of N, where they show distinct features. The curves for P are also presented for comparison. The main observation is that $P_1$ is always proportional to N, no matter where $P_{CO}$ is, and it is always much larger than $P_2$. The behavior of $P_2$ depends on the parameter $P_{CO}$: For $P_{CO} > HB$, see panels (c) to (f), $P_2$ also increases linearly with N; But for $P_{CO} < HB$, $P_2$ scales as $N^1$ when N is small and $N^0$ when N is large.

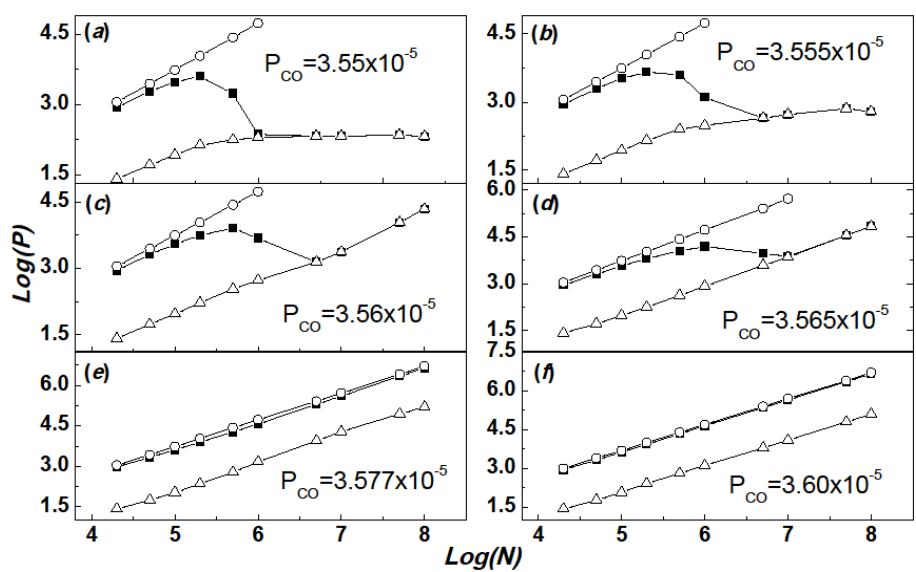

Fig. 6. The total entropy production $P$ (■), that of spikes $P_1$ (o), and of small oscillations $P_2$ (Δ) are plotted as functions of the system size N for different control parameter $P_{CO}$.

Combine above features, one may qualitatively understand the nontrivial behaviors of P shown in Fig.4 as follows,

(i) For $P_{CO} > CANARD$, the deterministic system shows spikes. In this region, both $P_1$ and $P_2$ are proportional to N, and $\alpha$ is nearly one. Thus $P \sim P_1$ and it is naturally that $P \propto N^1$.

(ii) For $P_{CO} < HB$, the deterministic system shows a steady state. When N is large, $\alpha$ is nearly zero and $P \sim P_2$, leading readily to $P \propto N^0$. When N is small, noting that $\alpha \gg 1 - \alpha$ and $P_1 \gg P_2$, we have $P \sim P_1$, which is in accordance with the linear relationship. In the intermediate N region, one may write $P_1 \sim c_1 N$, $P_2 \sim c_2 N$, $\alpha \sim e^{-bN}$, where $c_1 > c_2 > 0$ and $b > 0$ are constants.

Summing these terms together, it is easy to show that a maximum arises at some optimal value of *N*.

(iii) For $P_{CO}$ between the HB and the CANARD, the deterministic system shows small amplitude oscillation but no spikes. In this region, both $P_1$ and $P_2$ are also both proportional to N, which is similar to the case when $P_{CO}$ is larger than the CANARD. Thus in the large *N* limit, P is mainly contributed by $P_2$ and should be linearly dependent on *N*. The same reasoning as in (ii) for $P_{CO} < HB$ also holds here, which may qualitative illustrate the linear dependence in the small *N* range and the occurrence of extreme in the intermediate *N* range.

## 3. Conclusions

In summary, we have applied the concept and principles of ST to a mesoscopic chemical oscillation system with both oscillatory and excitable dynamics. The multiscale nature of the system and the existence of a CANARD lead to interesting noise-induced dynamics. Generally, a stochastic trajectory contains of a number of spikes lying on the background of small amplitude oscillations. We find that the time averaged entropy productions, contributed from the spikes or from the small oscillations, show quite different dependences on the system size. This also leads to nontrivial dependence of the total entropy production on the system size, even including a maximum for certain control parameters. This work thus unravels the very relationship between the noisy dynamics of a stochastic system and ST features. Since excitable systems with oscillatory dynamics are ubiquitous in many chemical, biological and physical disciplines, our study could open more perspectives regarding the application of ST.

**Acknowledgments**:This work is supported by the National Science Foundation of China (Grant No. 20873130, 20933006).